\documentclass[
aps,prl
 amsmath,amssymb,
 reprint,%
]{revtex4-1}
\usepackage{chngcntr} % \counter
\usepackage{etoolbox} % For \AtBeginEnvironment
\usepackage{graphicx}
\usepackage{float} 
\usepackage[mathlines]{lineno}% Enable numbering of text and display math
\usepackage[font=small]{caption,subcaption}
\usepackage{dcolumn}% Align table columns on decimal point
\usepackage{bm}% bold math
\usepackage{tikz}
\makeatletter
\@ifpackageloaded{float}{%
  \AtBeginEnvironment{figure}{\nolinenumbers}%
  \AtBeginEnvironment{figure*}{\nolinenumbers}%
  \AtBeginEnvironment{table}{\nolinenumbers}%
  \AtBeginEnvironment{table*}{\nolinenumbers}%
}{}
\makeatother
\usepackage{mathptmx}
\usepackage{color}
\usepackage[colorlinks,linkcolor=black]{hyperref}
\usepackage{physics}
\usepackage{ulem}

\newcommand{\tabC}[2]{\begin{minipage}[t]{#1}\centering #2\end{minipage}}
\newcommand{\tabSignalsText}[1]{\tabC{0.68\linewidth}{#1}}

\begin{document}

\title{A real-time disruption prediction and mitigation system for the EXL-50U spherical torus}

\author{
J.~P.~Zhou$^{1}$,
S.~F.~Liu$^{1,*}$,
J.~Q.~Cai$^{2,\dagger}$,
H.~Y.~Zhao$^{2}$,
J.~Li$^{2}$,
Y.~P.~Zhang$^{2}$,
D.~Guo$^{2}$,
C.~Wu$^{2}$,
A.~Wang$^{2}$,
H.~Y.~Li$^{2}$,
C.~Zhang$^{2}$,
Z.~Y.~Chen$^{3}$,
Y.~J.~Shi$^{2}$\\[4pt]
$^{1}$School of Physics, Nankai University, Tianjin 300071, China\\
$^{2}$ENN Science and Technology Development Co., Ltd, Langfang 065001, China\\
$^{3}$School of Electrical and Electronic Engineering, Huazhong University of Science and Technology, Wuhan 430074, China\\
$^{*}$lsfnku@nankai.edu.cn, $^{\dagger}$caijianqinga@enn.cn
}

\begin{abstract}
This work presents a real-time disruption prediction and mitigation system developed for high-current operations in the EXL-50U Spherical Torus. By leveraging Reflective Memory (RFM) technology, the system establishes a low-latency real-time data path, creating a fully integrated pipeline that synchronizes multi-channel diagnostic acquisition, online preprocessing, real-time inference, and Massive Gas Injection (MGI) triggering. At its core, a lightweight prediction model based on a Temporal Convolutional Network (TCN) with a channel attention mechanism extracts disruption precursor features while adaptively weighting the importance of different diagnostic channels. {Tested across discharges \#14036--\#14790, the system achieves a true positive rate of 82.4\% and a false positive rate of 16.5\%, with end-to-end latency below $1~\mathrm{ms}$ in online operation.} Mitigation experiments further show that the MGI system can supply the required gas inventory and trigger a rapid post-injection plasma response, supporting the operational requirements of EXL-50U and providing engineering guidance for future devices such as EHL-2. These results confirm the engineering feasibility of integrated real-time disruption control on EXL-50U, offering a robust basis for future research in higher-parameter fusion devices.

\quad

\noindent\textbf{Keywords:} Real-time Disruption Prediction, Temporal Convolutional Network, Reflective Memory, Massive Gas Injection, EXL-50U Spherical Torus
\end{abstract}

\maketitle

\section{Introduction}\label{sec:introduction}
Plasma disruptions are among the most destructive transient events in tokamak experiments, posing a critical challenge to the safe and stable operation of fusion devices. During a disruption, the plasma rapidly loses magnetic confinement, triggering a thermal quench followed by a current quench on a millisecond timescale. This process causes a rapid drop in plasma current and a sudden deposition of thermal energy, together with strong electromagnetic loading on plasma-facing components and surrounding structures. The resulting heat loads, electromagnetic forces, and potential runaway electron beams can cause severe damage to the first wall, divertor, and supporting structures, threatening the device's integrity and lifetime. As fusion research advances toward next-generation facilities like CFETR and ITER, where plasma energy and stored magnetic energy are significantly higher, the consequences of disruptions become even more severe. Reliable disruption prediction is therefore essential--not only to protect existing devices, but also to ensure the viability of future reactor-scale experiments. Without timely and accurate warning, mitigation systems cannot be triggered early enough to radiate thermal energy, mitigate concentrated heat loads, or suppress runaway electrons.

In recent decades, significant advancements have been made in disruption prediction methodologies across major tokamak facilities worldwide. The field has broadly evolved from physics-inspired threshold criteria toward increasingly sophisticated data-driven models, some of which have been implemented in real-time plasma control environments. Early efforts, such as those on JET and ASDEX Upgrade, established the foundational "prediction-mitigation" framework. These systems monitored parameters such as locked modes, radiation levels, and Greenwald density limits. When predefined thresholds were crossed, they would trigger actions like discharge termination or gas injection \cite{Moreno,LehnenNF}. Building on this foundation, JET further advanced data-driven disruption prediction through its Automatic Disruption Prediction System (APODIS). Deployed online during the ITER-like wall campaign, this multilayer support vector machine architecture represented a significant leap, demonstrating that machine learning could maintain a high success rate and low false alarm rate in a real-time environment \cite{Moreno}. {
DIII-D embedded its Random Forest-based algorithm, Disruption Prediction via Random Forest (DPRF), directly into the plasma control system, achieving warning times of several hundred milliseconds over more than 900 discharges; its ability to provide feature importance analysis offered new insights into disruption physics and control strategy design \cite{Rea}. EAST developed a real-time predictor tailored for high-density disruption scenarios. By testing its Random-Forest model in piggyback operation and dedicated control-system experiments, EAST provided valuable experience for deploying data-driven disruption predictors on superconducting tokamaks \cite{Hu}. On J-TEXT, an online hybrid neural-network system for density-limit disruption prediction has been coupled with the real-time density feedback control system; when a disruption is predicted, the gas puffing control valve is closed immediately to avoid density-limit disruptions, achieving an average warning time of about $40~\mathrm{ms}$ \cite{Zheng}. More recently, KSTAR has extended data-driven disruption prediction from database-based random-forest modeling to real-time integrated prediction and mitigation in around $1~\mathrm{MA}$ plasmas, demonstrating the continued progress of data-driven approaches toward practical deployment in high-performance tokamak operation \cite{LeeRF,LeeNF}.
}

Following the heating upgrade, EXL-50U has entered a phase of high-current operation, during which the risk of device damage caused by disruptions has increased significantly. To ensure safe operation, a real-time prediction and mitigation system with millisecond-level response is urgently required. This paper presents the development and experimental validation of a real-time disruption prediction and mitigation system designed for EXL-50U operation. The system establishes an integrated real-time processing chain that synchronizes multi-channel diagnostic acquisition, disruption prediction, and mitigation triggering, thereby enabling predictive results to be translated into executable control actions. Experimental results demonstrate that the system provides reliable warning signals under real operating conditions and successfully identifies most disruption events, confirming its engineering feasibility and practical applicability.

The remainder of this paper is organized as follows. Section II starts with an overview of the EXL-50U device, including its basic configuration, diagnostic systems, and the real-time framework. {Section III then details the disruption prediction algorithm based on the TCN-Attention model, covering dataset construction, model architecture, model comparison and ablation studies, training procedure, and experimental results.} Section IV describes the design and implementation of the disruption mitigation system. Finally, Section V concludes the paper with a summary of key findings and an outlook on future research directions.

\section{Experiment Setup and Device}\label{sec:experiment}

\subsection{EXL-50U Spherical Torus and Diagnostics}

EXL-50U is a spherical tokamak developed by ENN Science and Technology Development Company as an upgrade of its predecessor, EXL-50, and is regarded as China’s first large spherical torus device. The device features a compact integrated center-column structure, wherein the Central Solenoid (CS) is coaxially arranged with a portion of the Toroidal Field (TF) coils. Ten sets of Poloidal Field (PF) coils are installed outside the vacuum vessel, enabling flexible magnetic equilibrium configurations, including limiter, single-null divertor, and double-null divertor geometries. The key engineering parameters of EXL-50U are as follows: major radius of $0.6$--$0.8~\mathrm{m}$, toroidal magnetic field up to $1~\mathrm{T}$, aspect ratio of $1.4$--$1.85$, and elongation of $1.4$--$2$. The TF coils are capable of delivering a $1.2~\mathrm{T}$ flat-top field for a duration of $2.5~\mathrm{s}$ \cite{Shi1}.

In addition to its advanced magnetic configuration capabilities, EXL-50U is equipped with a comprehensive suite of auxiliary heating and current drive systems. These include one Neutral Beam Injection (NBI) system rated at $50~\mathrm{keV}/1.5~\mathrm{MW}/5~\mathrm{s}$, three $28~\mathrm{GHz}/400~\mathrm{kW}/5~\mathrm{s}$ Electron Cyclotron Resonance Heating (ECRH) systems, one $50~\mathrm{GHz}/400~\mathrm{kW}/5~\mathrm{s}$ ECRH system, one Lower Hybrid Current Drive (LHCD) system at $2.45~\mathrm{GHz}/200~\mathrm{kW}/5~\mathrm{s}$, and one Ion Cyclotron Resonance Heating (ICRH) system with a tunable frequency range of $3$--$26~\mathrm{MHz}$ and output power of $100~\mathrm{kW}/5~\mathrm{s}$ \cite{Shi1}. To support high-performance plasma operations, the device integrates a robust plasma control framework. With the assistance of RZIP closed-loop shape control, a magnetic probe system, and the EFIT/CCS real-time equilibrium reconstruction code, EXL-50U achieves precise regulation of plasma configurations, including shape and position control. Furthermore, the device features an extensive diagnostic suite essential for both physical studies and real-time control. Key systems include Thomson scattering (TS) \cite{LiTS}, an AXUV array \cite{C. Zhang}, a soft X-ray array \cite{HuangSXR}, an HCN interferometer \cite{XieInterferometer}, magnetic probes, a fiber optic current sensor (FOCS) \cite{LiEddy}, a hard X-ray spectrometer \cite{Qi,ChengHXR}, as well as visible and infrared imaging diagnostics \cite{GuoIRVisible,GuoOptical}. Collectively, these diagnostics provide rich, high-resolution datasets that are critical for plasma parameter characterization and serve as the foundation for data-driven disruption prediction research.

EXL-50U achieved its first plasma discharge on January 19, 2024, with an initial target plasma current of approximately $500~\mathrm{kA}$ \cite{Shi1}. Through subsequent hardware optimization and accumulated operational experience, the discharge performance has steadily improved. In April 2025, EXL-50U successfully demonstrated a $1~\mathrm{MA}$-level hydrogen-boron plasma discharge, marking a significant milestone in its capability to operate at high current and high energy density \cite{Shi2}. However, the increase in operational parameters has also intensified magnetohydrodynamic (MHD) activity and elevated the risk of plasma disruptions. Consequently, under the current high-performance operating regime, the development of an online disruption prediction and mitigation system capable of millisecond-level response has become essential to ensure the safe and reliable operation of the device \cite{Shi1}.

\begin{figure*}[t]
\centering
\includegraphics[width=0.95\textwidth]{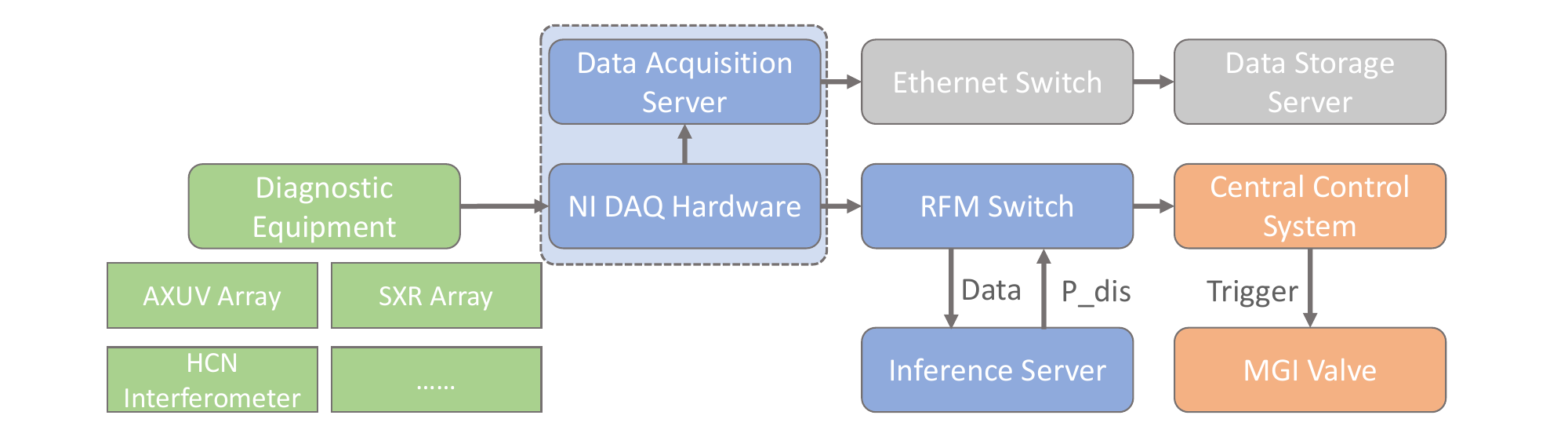}
\caption{Architecture of the real-time disruption prediction and mitigation system based on RFM network.}
\label{Architecture}
\end{figure*}

\subsection{Real-time Module}

{
To satisfy the real-time requirements of disruption prediction, a distributed data acquisition and processing module based on Reflective Memory technology has been deployed on EXL-50U to support online prediction and mitigation experiments. The module utilizes the GE-5565 RFM card as the core component for high-speed real-time communication. An optical fiber network interconnects multiple nodes to form a deterministic RFM network with predictable latency. When a node writes data to its local RFM memory, the onboard hardware automatically broadcasts the data and synchronizes it to the corresponding memory addresses of all other nodes, enabling shared-memory-style data exchange with low latency and high reliability. The GE-5565 provides $128~\mathrm{MB}$ of memory and supports multi-node access \cite{RFM1,RFM2}.

\begin{figure}[t]
\centering
\includegraphics[width=\linewidth]{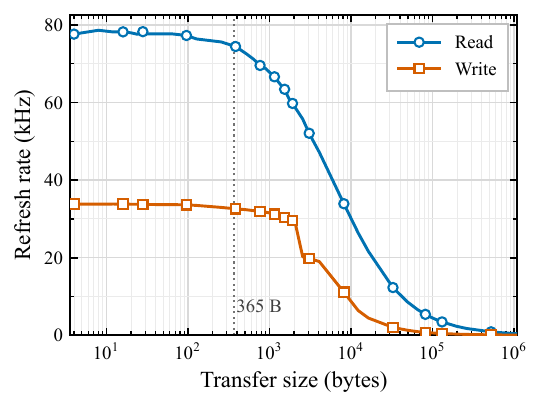}
\caption{GE-5565 RFM read and write refresh rates versus transfer size from the RFM performance test (reported as IOps for sustained read/write at each block size, i.e., the inverse of the average transfer time). The vertical dashed line at 365~B marks the transfer size of each write from the data acquisition server to the RFM on EXL-50U.}
\label{fig:rfm_refresh}
\end{figure}

Figure~\ref{fig:rfm_refresh} shows the RFM read and write refresh rates as a function of transfer size. In the EXL-50U real-time link, the data acquisition server writes 365-byte diagnostic data to the RFM at a refresh rate of 10~kHz. Because both model training and online inference use a uniform data rate of 1~kHz, the inference server reads the preprocessed channel data required by the model from the RFM every 1~ms. The write refresh rate at 365~B is well above the 10~kHz requirement, and the read refresh rate is also well above the online inference rate, fully meeting the stringent demands of the real-time disruption prediction system for fast and deterministic data transfer.

The architecture of the real-time disruption prediction system based on the GE-5565 RFM module is illustrated in Figure~\ref{Architecture}. The system adopts a multi-node distributed design and primarily comprises a data acquisition server, a data storage server, an inference server, the Central Control System (CCS), and National Instruments (NI) data acquisition hardware installed in the data acquisition server. The NI data acquisition hardware acquires multi-channel diagnostic signals from EXL-50U at a sampling rate of 200~kHz with a resolution of 16~bit; all channels required for disruption prediction are processed using the same configuration. Multi-channel timing synchronization is implemented using a PXI-6683H timing and synchronization module. This module uses GPS signals as an external time reference to calibrate the system clock, and all server nodes are aligned to this reference, thereby achieving nanosecond-level synchronization among the nodes. The raw diagnostic data acquired by the NI data acquisition hardware are subsequently divided into two processing paths. In the data storage path, the diagnostic data are transmitted to the data acquisition server via optical fiber and averaged over every five samples to generate 40~kHz data. These data are subsequently transmitted via Ethernet to the data storage server for offline analysis and model training. In the real-time processing path, every twenty samples are averaged to generate 10~kHz data, which are subjected to real-time preprocessing and then written to a designated address space in the RFM.
The inference server reads the preprocessed data from the RFM every 1~ms and performs real-time inference using the TCN-Attention model, which is deployed with TensorRT acceleration, to output the predicted disruption probability. This prediction result is subsequently written back to the RFM and read in real time by the CCS for threshold evaluation. When the predicted probability exceeds a preset threshold, the system issues a Massive Gas Injection (MGI) command to initiate disruption mitigation.

}

\section{Prediction Module}\label{sec:prediction}

{
This section presents the disruption prediction module developed for real-time operation on EXL-50U. The discussion begins with the construction of the disruption prediction dataset, followed by a detailed description of the proposed prediction model and model comparison and ablation studies. The training and inference performance of the model are then evaluated, with particular emphasis on both predictive accuracy and the system's real-time capability. The results demonstrate that the proposed module can effectively identify disruption precursors while satisfying the stringent latency requirements for real-time warning and subsequent mitigation on EXL-50U.
}

\subsection{Datasets}

{
For the construction of the disruption prediction database, the selection of diagnostic channels must satisfy two essential criteria. First, the signals should contain physically meaningful information indicative of disruption precursors; second, these signals must be accessible within the real-time data processing chain. Regarding physical relevance, we surveyed the literature to identify diagnostics strongly associated with disruptions and selected the corresponding channels on EXL-50U with adequate signal quality \cite{Aymerich,Shen,YangHL2A,YangHL3,Church}. Regarding real-time availability, prior to this work the real-time data link primarily served plasma control; with the commencement of disruption-prediction work, radiative and density channels relevant to disruption precursors were further integrated into the link. At that stage, parameters from PTEFIT (PyTorch-TensorRT-EFIT) equilibrium reconstruction \cite{PTEFIT} (such as $q_{95}$ and internal inductance) and locked-mode signals from saddle coils were not yet available in real time and were therefore not used as online inference inputs. Based on these two principles, eight diagnostic channels were ultimately selected from the EXL-50U diagnostic system as model inputs.
} These signals encompass several key physical quantities associated with disruption precursors, including radiated power (AXUV), electron density (HCN), plasma current (IP), and plasma position (ZP and RP). All selected signals are acquired by the high-speed data acquisition system and transmitted via the RFM network, serving as inputs for real-time inference. Table~\ref{tab:signals} lists the selected channels and their physical interpretations.

\begin{table}[t]
\centering
\caption{Model input signal channels}
\label{tab:signals}
\footnotesize
\setlength{\tabcolsep}{4pt}
\begin{tabular*}{\columnwidth}{@{\extracolsep{\fill}}c@{\hspace{5pt}}c@{}}
\hline
Channel & \tabSignalsText{Physical meaning} \\
\hline
AXUV001 & \tabSignalsText{Poloidal AXUV radiation signal (lower boundary)} \\
AXUV016 & \tabSignalsText{Poloidal AXUV radiation signal (core)} \\
CCIP    & \tabSignalsText{Reference current} \\
IP      & \tabSignalsText{Plasma current} \\
I\_TF   & \tabSignalsText{Toroidal field current} \\
HCN\_NE001 & \tabSignalsText{Core chord average density} \\
ZP      & \tabSignalsText{Vertical displacement} \\
RP      & \tabSignalsText{Horizontal displacement} \\
\hline
\end{tabular*}
\end{table}

The discharges utilized in this study were selected from the range of shots \#10226 to \#14035. To ensure sample quality and label reliability, the raw discharge data underwent a screening process. Discharges with excessively low plasma current (below $150~\mathrm{kA}$) and those manually marked as invalid were excluded from further analysis. After screening, a total of 802 shots were retained for model training and validation, {and all channel data for these shots are stored at a 1~kHz sampling rate}. It should be noted that the model input does not treat each full discharge as a single sample. Instead, multiple fixed-length slices are extracted from the discharge time series using a sliding time window. This approach ensures compatibility with the model input format and captures the local temporal evolution at different time instants. In this work, a sliding window of $20~\mathrm{ms}$ with a step size of $5~\mathrm{ms}$ was adopted to segment the original discharge sequences. Combined with the selected diagnostic channels, the resulting dataset is structured as a three-dimensional tensor with dimensions corresponding to the number of samples, the number of diagnostic channels, and the time window length. Table~\ref{tab:dataset_composition} presents the composition of the dataset in terms of both discharges and slices, illustrating the data distribution at the pulse level and the sample level, respectively.

\begin{table}[t]
\centering
\caption{Composition of the training and validation datasets.}
\label{tab:dataset_composition}
\begin{tabular*}{\columnwidth}{@{\extracolsep{\fill}}ccccc@{}}
\hline
Dataset & \multicolumn{2}{c}{Disruption} & \multicolumn{2}{c}{Regular pulses} \\
\hline
 & Pulses & Slices & Pulses & Slices \\
Training   & 384 & 9334  & 205 & 37631 \\
Validation & 153 & 3744  & 60  & 10515 \\
Total      & 537 & 13078 & 265 & 48146 \\
\hline
\end{tabular*}
\end{table}

\begin{figure*}[t]
\centering
\includegraphics[width=0.88\textwidth]{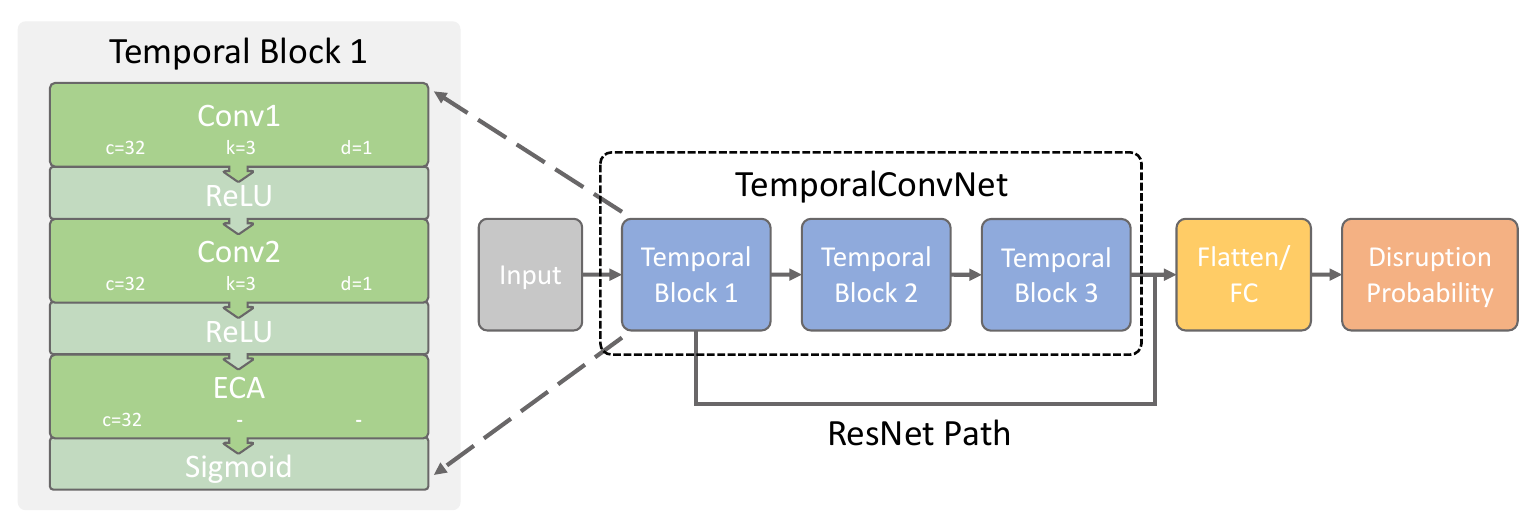}
\caption{Architecture of the TCN-Attention disruption prediction model. The model consists of three TemporalBlocks, each containing two one-dimensional convolutional layers and a channel attention module (ECA).}
\label{Architecture2}
\end{figure*}

\subsection{Model}

In this work, a Temporal Convolutional Network~(TCN)~\cite{Bai} is adopted as the backbone architecture, and an Efficient Channel Attention (ECA) mechanism~\cite{HuSE,Wang} is integrated to construct the TCN-Attention disruption prediction model. The overall architecture of the proposed model is shown in Figure~\ref{Architecture2}. It consists of three temporal convolution blocks (TemporalBlock) stacked in sequence, each containing two one-dimensional convolutional layers and a channel attention module.

The TCN constructs a large temporal receptive field by stacking dilated causal one-dimensional convolutional layers, enabling it to capture long-range temporal dependencies while maintaining computational efficiency. Given an input sequence $\mathbf{X} \in \mathbb{R}^{T \times C}$, the dilated causal convolution along the temporal dimension can be formulated as
\begin{equation}
    (\mathbf{X}*_d\mathbf{W})(t)=\sum_{i=0}^{k-1}{\mathbf{W}(i)\mathbf{X}(t-d\cdot i)}.
    \label{eq2}
\end{equation}
Here, $k$ denotes the kernel size and $d$ is the dilation factor~\cite{Oord}. The causal constraint ensures that the output at time step $t$ depends only on current and past inputs, thereby preventing future information leakage. Compared to recurrent architectures, TCN offers better parallelism and more stable gradient propagation, making it well suited for real-time prediction tasks.

To adaptively adjust the contribution of different diagnostic signals along the channel dimension, an Efficient Channel Attention module is incorporated after each TemporalBlock in the proposed model. The ECA module first generates a channel-wise descriptor via global average pooling, and then captures inter-channel dependencies using a local one-dimensional convolution:
\begin{equation}
    \boldsymbol{\omega }=\sigma (\mathrm{Conv1D}_k(\mathbf{g})).
    \label{eq3}
\end{equation}
Here, $\mathbf{g}$ denotes the channel-wise aggregated descriptor and $\boldsymbol{\omega}$ represents the learned channel attention weights~\cite{HuSE,Wang}. The kernel size $k$ is adaptively determined based on the channel dimension, enabling efficient attention modeling without introducing dimensionality reduction. The resulting weights are applied to recalibrate the temporal features, thereby enhancing informative channels while suppressing irrelevant or noisy ones.

In the three TemporalBlocks, the convolution kernel size is uniformly set to $k=3$, with dilation factors of $1$, $2$, and $4$, respectively. This configuration helps ensure that the effective temporal receptive field covers the key timescales relevant to disruption precursor evolution. Each convolutional layer is followed by a ReLU activation and equipped with a residual connection to improve training stability. A summary of the complete network architecture is provided in Table~\ref{tab:network_structure}.

\begin{table}[t]
\centering
\caption{Detailed network structure of the TCN-Attention model.}
\label{tab:network_structure}
\footnotesize
\setlength{\tabcolsep}{2pt}
\begin{tabular*}{\columnwidth}{@{\extracolsep{\fill}}c@{\hspace{1pt}}ccccc@{}}
\hline
Block & Layer & Kernel number & Kernel size & Dilation & Activation \\
\hline
\tabC{0.22\linewidth}{TemporalBlock1} & Conv1 & 32 & 3 & 1 & ReLU \\
 & Conv2 & 32 & 3 & 1 & ReLU \\
 & ECA   & 32 & -- & -- & Sigmoid \\
\tabC{0.22\linewidth}{TemporalBlock2} & Conv1 & 64 & 3 & 2 & ReLU \\
 & Conv2 & 64 & 3 & 2 & ReLU \\
 & ECA   & 64 & -- & -- & Sigmoid \\
\tabC{0.22\linewidth}{TemporalBlock3} & Conv1 & 128 & 3 & 4 & ReLU \\
 & Conv2 & 128 & 3 & 4 & ReLU \\
 & ECA   & 128 & -- & -- & Sigmoid \\
\hline
\end{tabular*}
\end{table}

{
\subsection{Model Comparison and Ablation Studies}

\begin{figure}[t]
\centering
\includegraphics[width=\linewidth]{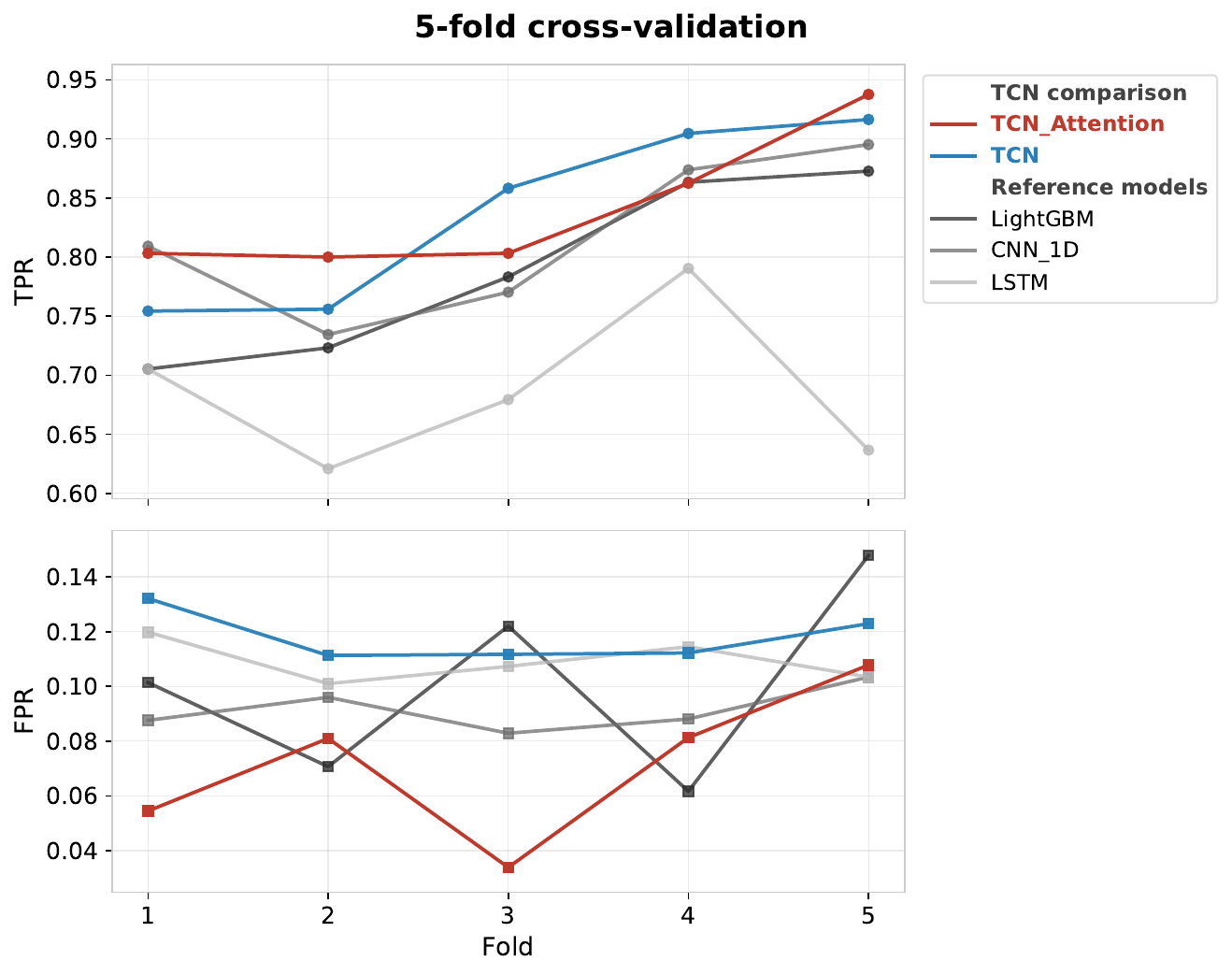}
\caption{Five-fold cross-validation results for model comparison and ablation studies. (a) True positive rate (TPR, i.e., recall) and (b) false positive rate (FPR) for each model.}
\label{fig:model_comparison}
\end{figure}

Because plasma operating conditions in EXL-50U experimental discharges vary across different periods, dividing the training and test sets by fixed time windows alone cannot fairly and reliably compare the performance of the different models. Therefore, we randomly shuffled the shot list and performed shot-level 5-fold cross-validation on the dataset described above, so that each fold contained disruptive and non-disruptive shots from different operating periods. In addition to the proposed TCN-Attention model, LightGBM, a one-dimensional CNN, and LSTM were selected as reference models; a TCN model without the ECA module was also included to ablate the contribution of channel attention. Figure~\ref{fig:model_comparison} presents the 5-fold cross-validation results: the upper panel shows the true positive rate (TPR, i.e., recall) for each model, and the lower panel shows the false positive rate (FPR). It should be noted that, compared with fixed time-window splits, the cross-validation metrics are generally more favorable; this is because randomizing the shot list mixes discharges from different operating periods within each test fold, so that the training and test sets are more similar in operating-condition distribution. TCN-Attention achieved an average TPR of 84.1\% and an average FPR of 7.2\% across the five folds, delivering the best overall performance; as can be seen in the figure, the TCN model attains a TPR comparable to TCN-Attention, whereas incorporating the ECA module markedly reduces the false alarm rate. Based on these results, TCN-Attention was selected as the online deployment model for the EXL-50U real-time disruption prediction system.
}

\subsection{Training and Inference Performance Evaluation}

\begin{figure*}[t]
\centering
\includegraphics[width=0.8\linewidth]{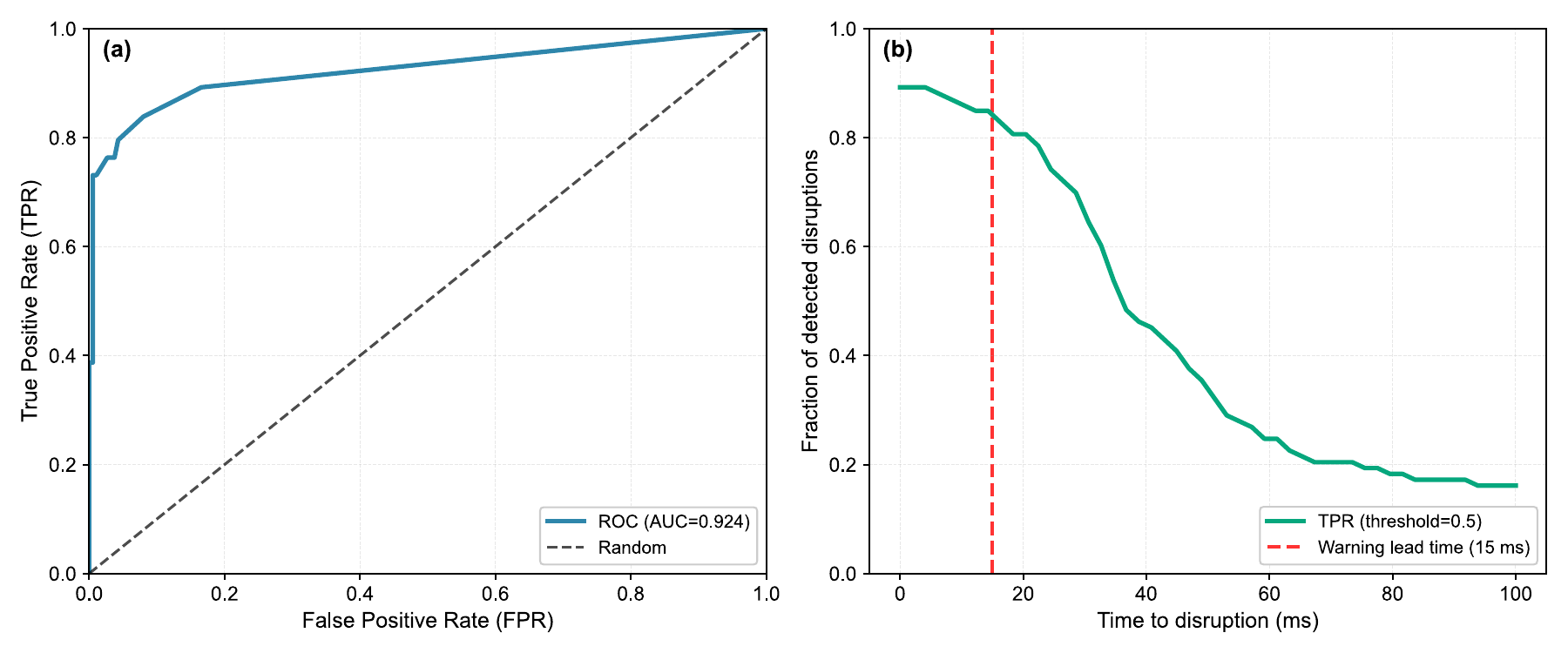}
\caption{Performance evaluation of the proposed disruption prediction system on the validation set. (a) ROC curve at a warning lead time of $15~\mathrm{ms}$. (b) Fraction of detected disruptions versus warning lead time.}
\label{Performance}
\end{figure*}

The model was implemented using the PyTorch deep learning framework and trained on a single NVIDIA GeForce RTX 4060 GPU. Network weights were initialized using Xavier initialization, and the Adam optimizer was employed for parameter updates. The training process lasted for 18 epochs, after which the model converged. The trained model was first evaluated on the validation set. {
By scanning different warning thresholds and required lead times, a practical real-time warning rule for online deployment was determined. Based on this analysis, a disruption warning is triggered once the predicted disruption probability exceeds a threshold of $0.5$.
} Under this rule, the performance metrics shown in Figure~\ref{Performance} were obtained. Figure~\ref{Performance}(b) shows how the fraction of disruptive shots correctly detected in time varies with the required warning lead time. As the required lead time increases, this fraction generally decreases, indicating that maintaining reliable early warnings becomes more difficult when more advance time is demanded. Notably, when the warning lead time is set to $15~\mathrm{ms}$—a duration sufficient for the disruption mitigation system to respond—the corresponding ROC curve shown in Figure~\ref{Performance}(a) yields an AUC of $0.924$. This result demonstrates that the model maintains strong discriminative capability even under the constraint of a practically required lead time.

To further validate the proposed system, experimental evaluation was conducted on EXL-50U using a test set comprising 160 discharges (shots \#14036--\#14790). The warning rule determined from the validation analysis was directly applied to the test set, and a prediction was considered successful only if the warning was issued at least $15~\mathrm{ms}$ before the actual disruption. The confusion matrix for the online test is shown in Table~\ref{tab:confusion_matrix}. Among the 160 discharges, the system correctly identified 42 disruptive events (True Positives, TP = 42) and 91 non-disruptive discharges (True Negatives, TN = 91). It falsely classified 18 non-disruptive discharges as disruptions (False Positives, FP = 18) and missed 9 disruptive events (False Negatives, FN = 9). {
These results yield a TPR of 82.4\% and an FPR of 16.5\%. The deployed TCN-Attention model identifies most disruption events under real operating conditions, and successful predictions provide a warning time window sufficient to support subsequent mitigation actions.
}

\begin{table}[t]
\centering
\caption{Confusion matrix of the real-time online prediction results for shots \#14036--\#14790.}
\label{tab:confusion_matrix}
\begin{tabular*}{\columnwidth}{@{\extracolsep{\fill}}ccc@{}}
\hline
 & Predicted Disruption & Predicted Normal \\
\hline
Actual Disruption & TP = 42 & FN = 9 \\
Actual Normal     & FP = 18 & TN = 91 \\
\hline
\end{tabular*}
\end{table}

\subsection{Latency Analysis}

\begin{figure}[t]
\centering
\includegraphics[width=\linewidth]{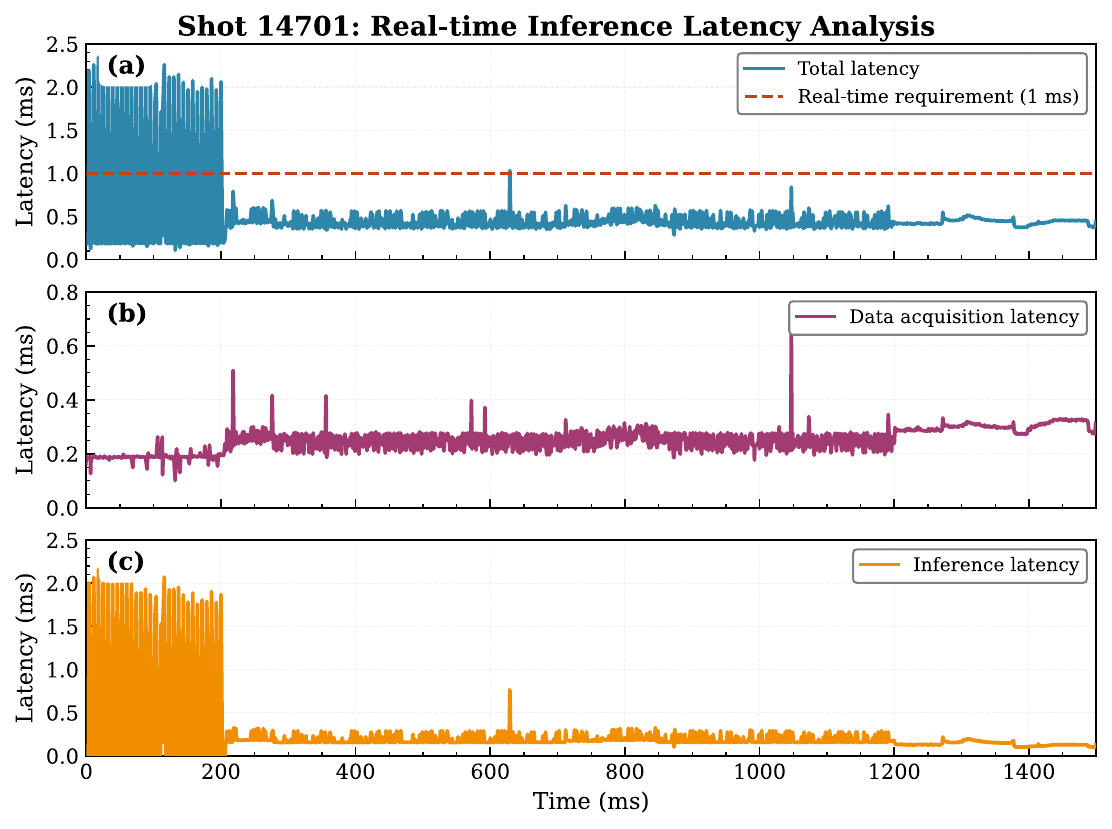}
\caption{Time series of system latency during a typical discharge (shot \#14701): (a) model inference latency, (b) data acquisition and transmission latency, and (c) end-to-end system latency.}
\label{fig:latency_series}
\end{figure}

A disruption prediction and mitigation system must not only accurately distinguish between disruptive and non-disruptive states but also satisfy stringent real-time requirements to enable timely warnings and responses upon detecting disruption precursors. To evaluate this capability, the end-to-end latency of the proposed system was measured under real operating conditions on EXL-50U. The assessment encompassed the entire processing chain, including multi-channel signal acquisition, data preprocessing, model inference, and control command generation. Figure~\ref{fig:latency_series}(c) shows the time series of the system's end-to-end latency during online operation for shot \#14701. The inference latency is relatively high during the first approximately $200~\mathrm{ms}$ after system startup. This behavior is primarily attributed to the initialization and optimization procedures of the TensorRT inference engine. This issue can be effectively mitigated by performing model warm-up before formal experimental operation. After warm-up, the inference latency stabilizes consistently within the range of $0.2$--$0.3~\mathrm{ms}$.

{
As shown in Figure~\ref{fig:latency_series}(b), the data acquisition and transmission latency remains within the range of $0.2$--$0.3~\mathrm{ms}$, demonstrating that the RFM-based real-time data link provides stable low-latency communication. To further characterize the distribution of total end-to-end latency, 1000 valid measurement points were extracted from each of ten online discharges: after model warm-up, samples were taken every $1~\mathrm{ms}$ over the interval from $200$ to $1200~\mathrm{ms}$, yielding 10,000 samples in total.

\begin{figure}[t]
\centering
\includegraphics[width=\linewidth]{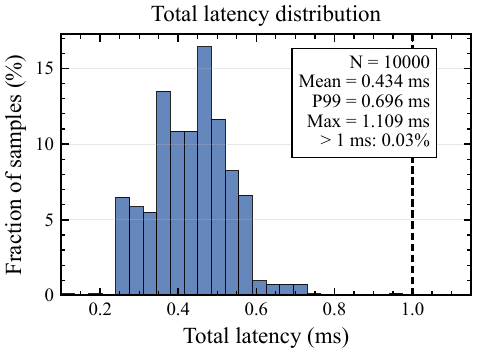}
\caption{Histogram of total end-to-end system latency based on 10,000 valid measurements from ten online discharges (sampled every $1~\mathrm{ms}$ from $200$ to $1200~\mathrm{ms}$ after model warm-up).}
\label{fig:e2e_latency_hist}
\end{figure}

The histogram of total end-to-end latency is shown in Figure~\ref{fig:e2e_latency_hist}; across all samples, only three time points exhibit latency slightly above $1~\mathrm{ms}$. This duration is shorter than the characteristic timescales of tokamak disruptions. In present-day tokamaks, the thermal quench typically occurs on a timescale of about $1~\mathrm{ms}$, whereas the subsequent current quench can last from a few to several hundred milliseconds \cite{Vega,Boozer}. These results confirm that the selected diagnostic inputs, the RFM-based real-time transmission link, and the online inference architecture collectively satisfy the timing requirements for real-time disruption prediction and subsequent mitigation triggering on EXL-50U.
}

\section{Mitigation Module}\label{sec:mitigation}

To establish a complete prediction-response chain for safe device operation, the system must rapidly trigger the mitigation actuator once a disruption warning is issued. Therefore, real-time prediction alone is insufficient; a mitigation module with adequate response speed and actuation capability is also essential for effective plasma intervention. To meet this requirement, this work evaluates the engineering applicability of the disruption mitigation module on EXL-50U. Once a warning is generated by the prediction system, the module promptly sends a trigger signal to the central control system and drives the mitigation actuator to respond rapidly, thereby reducing the thermal loads, electromagnetic forces, and associated operational risks during a disruption. In this study, MGI was selected as the mitigation actuator. The following sections describe its mitigation mechanism, hardware implementation, and trigger response characteristics.

\subsection{MGI for Disruption Mitigation}

Massive Gas Injection is a widely adopted disruption mitigation technique in tokamak research. Its fundamental principle involves the rapid injection of a large quantity of neutral gas--such as deuterium, neon, or argon--into the plasma upon detection of disruption precursors or fulfillment of trigger conditions, aiming to achieve controlled dissipation of plasma thermal energy and current \cite{LehnenMGI}. Upon MGI triggering, the injected gas becomes rapidly ionized and significantly enhances volumetric radiation losses through ionization and excitation processes. As a result, the plasma thermal energy is radiated over an extremely short timescale, accelerating the thermal quench (TQ) and effectively mitigating the peak heat flux deposited on the first wall and divertor \cite{LehnenMGI,Pautasso}. Concurrently, the introduction of impurities leads to a sharp increase in the effective plasma resistivity, facilitating faster dissipation of the plasma current during the current quench (CQ) phase. This accelerated current decay helps reduce the electromagnetic forces and halo current loads that pose risks to the device structure \cite{Pautasso,LehnenJNM}.

In addition to mitigating thermal and electromagnetic loads, MGI can also help suppress runaway electron generation during disruptions by increasing plasma density and enhancing collisional damping \cite{Hollmann}. The efficacy of this approach has been demonstrated in large tokamaks such as JET, where MGI systems have been shown to reduce disruption-induced thermal loads, electromagnetic forces, and runaway electron risks, thereby helping to protect plasma-facing components \cite{LehnenMGI,Hollmann}. Building on this mature experimental foundation, MGI has become a well-established disruption mitigation approach with strong engineering relevance for next-step high-power fusion devices.

\subsection{MGI Design Specifications}

\begin{figure*}[t]
\centering
\includegraphics[width=0.8\linewidth]{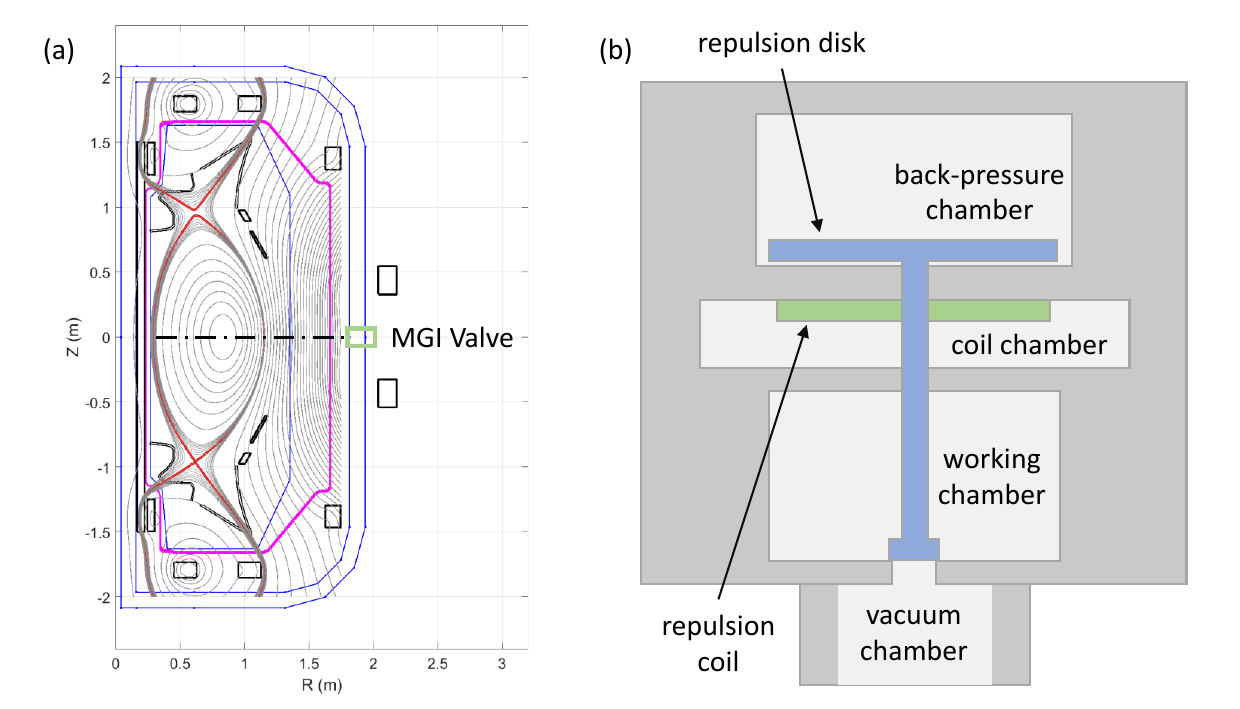}
\caption{Conceptual illustration of the MGI system in EXL-50U: (a) poloidal location of the MGI valve; (b) conceptual design of the MGI valve.}
\label{fig:mgi_structure}
\end{figure*}

As illustrated in Figure~\ref{fig:mgi_structure}(a), the MGI system in EXL-50U is installed at the low-field-side midplane and injects gas toward the plasma core with a radial injection angle of $0^\circ$. The detailed structure of the MGI valve is shown in Figure~\ref{fig:mgi_structure}(b), which consists of three main chambers: the back-pressure chamber, the coil chamber, and the working chamber. The sealing force of the valve is provided by the pressure difference between the high-pressure back-pressure chamber and the vacuum chamber. In standard operation, the back-pressure chamber is typically filled with gas at a pressure of 1--5~$\mathrm{bar}$. The coil chamber, which is separated from the back-pressure chamber by a stainless-steel layer, contains the electromagnetic repulsion coil and is open to the atmosphere. When triggered by a current pulse, the repulsion coil generates a transient electromagnetic force on the repulsion disk, causing the valve core to lift and open the flow path. As a result, the high-pressure gas stored in the working chamber is rapidly released into the vacuum chamber, thereby enabling massive gas injection. After the actuation pulse ends, the valve core returns to its initial position and reseals the valve, completing one operating cycle.

Based on this valve design, the key specifications of the MGI system are summarized below. The valve opens fully within $0.5~\mathrm{ms}$, as measured by an optical grating sensor, thereby meeting and surpassing the $1~\mathrm{ms}$ design specification required for timely disruption mitigation. With an injection capacity of up to $0.5 \times 10^{22}$ neon atoms, the system is designed to handle the heat flux mitigation requirements of EXL-50U. For comparison, approximately $10^{22}$ neon atoms have been predicted to be sufficient for full mitigation in the future EHL-2 device, a value comparable to reported injection quantities in existing tokamaks \cite{Cai}. Thus, the present system provides a robust solution for EXL-50U and offers useful engineering guidance for future MGI scaling toward devices such as EHL-2.

\subsection{TQ/CQ Response Delay After MGI Triggering}

As the actuator of the real-time disruption mitigation chain developed in this work, the engineering applicability of the MGI system hinges on two critical factors: (1) the ability to deliver a stable and controllable gas quantity under actual experimental conditions, and (2) the capability to induce a sufficiently fast plasma response following system triggering. To validate these aspects, the MGI system on EXL-50U was experimentally evaluated through both gas injection calibration and analysis of post-trigger plasma evolution.

To characterize the gas injection performance of the MGI system and identify suitable operating parameters, a systematic scan experiment was conducted under varying driving voltages. The experiments were performed in a strong magnetic field environment with all coils energized, thereby closely replicating actual operating conditions. The MGI valve was operated with a constant back pressure of $10~\mathrm{bar}$, and the injected gas quantity was regulated by adjusting the driving voltage across a range from $900~\mathrm{V}$ to $1500~\mathrm{V}$, comprising eight test points. The injected particle count was estimated based on pressure measurements from a vacuum gauge (GAS\_PRES02) located at the bottom of the vacuum vessel. Given the known vessel volume of $27~\mathrm{m}^{3}$, the number of injected particles corresponding to each driving voltage was calculated using the ideal gas law.

\begin{figure}[t]
\centering
\includegraphics[width=0.9\linewidth]{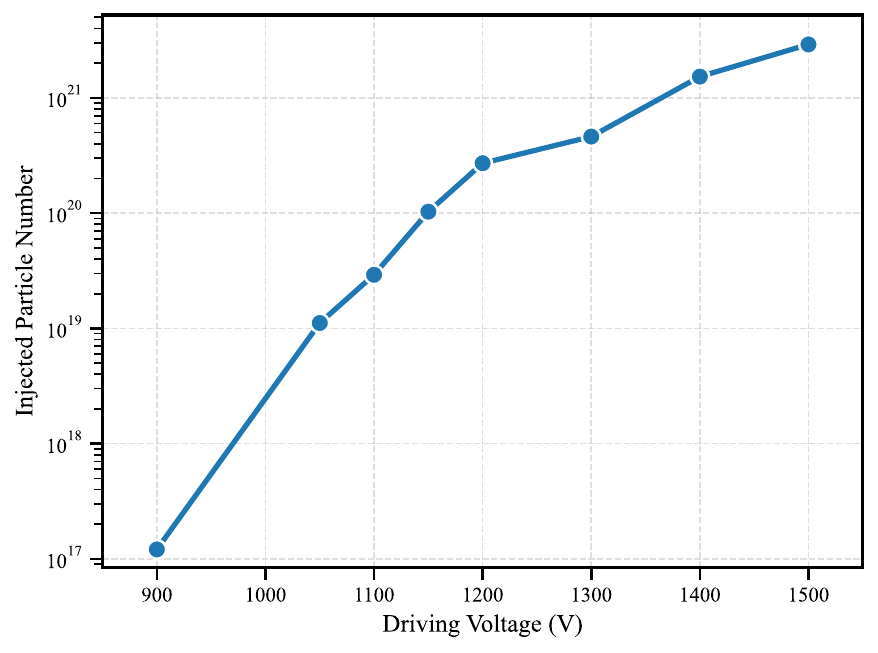}
\caption{Dependence of injected particle number and maximum pressure on driving voltage for the MGI system.}
\label{fig:mgi_calibration}
\end{figure}

As shown in Figure~\ref{fig:mgi_calibration}, the MGI system demonstrates stable and controllable gas injection performance across the tested driving voltage range. The injected particle number exhibits a strong positive dependence on the driving voltage, reaching a maximum value exceeding $2 \times 10^{21}$ at $1500~\mathrm{V}$. This monotonic relationship confirms the good parameter tunability of the system. The quantitative characterization presented here establishes a reliable foundation for optimizing the trigger strategy and determining the appropriate gas injection amount in subsequent disruption mitigation experiments.

\begin{figure}[t]
\centering
\includegraphics[width=\linewidth]{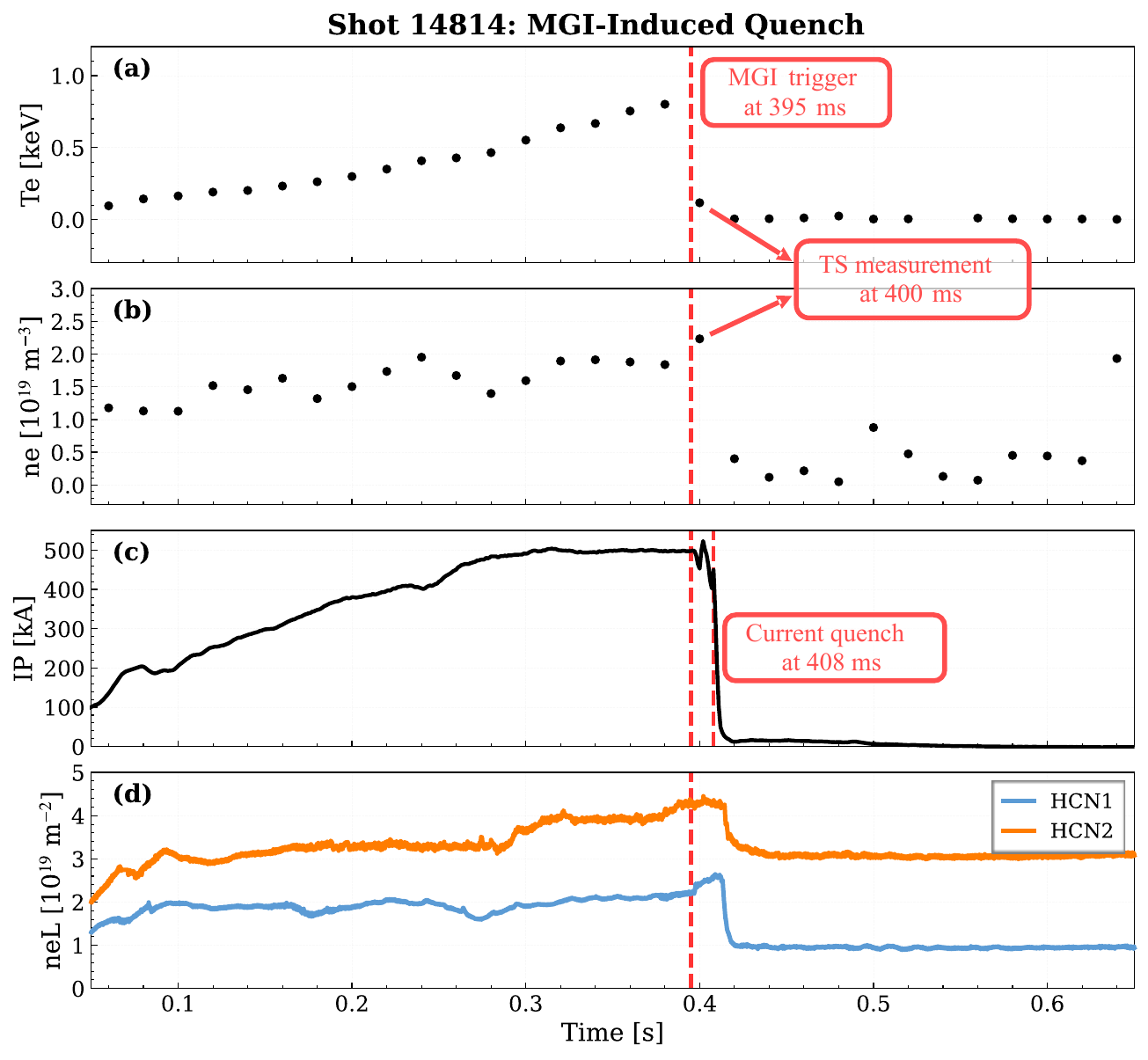}
\caption{Plasma response to MGI triggering at $t = 395~\mathrm{ms}$ for shot \#14814. (a) Core electron temperature (Thomson scattering). (b) Core electron density (Thomson scattering). (c) Plasma current. (d) Line-integrated electron density (HCN interferometer, $337~\mu\text{m}$).}
\label{fig:mgi_plasma_response}
\end{figure}

To assess the actual mitigation response of the MGI system, the time evolution of key plasma parameters was examined for shot \#14814, as shown in Figure~\ref{fig:mgi_plasma_response}. The plotted traces include core electron temperature and core electron density (both from Thomson scattering, with a sampling interval of $20~\mathrm{ms}$), plasma current, and line-integrated density from two HCN interferometer channels. The onset of the thermal quench is marked by a sudden collapse in the core electron temperature, while the current quench is evidenced by a rapid decay in the plasma current.

Prior to injection (around $380~\mathrm{ms}$), the core electron temperature remained stable at approximately $800~\mathrm{eV}$ (Figure~\ref{fig:mgi_plasma_response}(a)). Following gas injection, a sharp temperature collapse to about $150~\mathrm{eV}$ occurred at around $400~\mathrm{ms}$, corresponding to the onset of thermal quench approximately $5~\mathrm{ms}$ after MGI triggering. The current quench phase subsequently initiated at around $408~\mathrm{ms}$, as evidenced by the rapid decay of plasma current shown in Figure~\ref{fig:mgi_plasma_response}(c), approximately $13~\mathrm{ms}$ post-trigger. This sequence reflects the underlying physical mechanisms: impurity injection enhances radiative cooling, leading to thermal energy dissipation, while the concomitant increase in effective plasma resistivity accelerates current decay.

Collectively, the characterization of gas injection performance (Figure~\ref{fig:mgi_calibration}) and plasma response dynamics (Figure~\ref{fig:mgi_plasma_response}) validates that the MGI actuator on EXL-50U possesses the necessary response capability for real-time disruption mitigation. The actuator delivers controllable gas quantities and triggers a rapid thermal quench followed by current quench on millisecond timescales. When integrated with the real-time prediction chain demonstrated in previous sections, the complete system forms an effective and coordinated pipeline encompassing data acquisition, prediction, and mitigation. This integrated architecture indicates that, given a sufficient warning lead time, a practical engineering time window can be provided for disruption mitigation on EXL-50U.

\section{Summary}\label{sec:summary}

This work presents the design and implementation of a complete real-time disruption prediction and mitigation system for the EXL-50U tokamak. The system adopts a distributed architecture based on RFM technology, establishing a fully integrated real-time processing chain that encompasses multi-channel diagnostic signal acquisition, deep-learning-based model inference, and MGI trigger command generation. {The end-to-end latency is below $1~\mathrm{ms}$ under online operation, satisfying the stringent timing requirements for real-time disruption warning and mitigation triggering, thereby demonstrating the feasibility of deploying such a system on an operational device.}

For the prediction module, a lightweight disruption prediction model based on a Temporal Convolutional Network with channel attention (TCN-Attention) was developed. Evaluated on discharges spanning shots \#14036--\#14790, the model achieves a true positive rate (TPR) of 82.4\% and a false positive rate (FPR) of 16.5\%. These results indicate that the model can reliably identify the majority of disruption events under real operating conditions while providing a sufficient warning time window for subsequent mitigation actions. For the mitigation module, the actuation performance of the MGI system on EXL-50U was experimentally validated. The results demonstrate that the system enables controllable gas injection, with a maximum injected particle count exceeding $2 \times 10^{21}$. Following MGI triggering, a thermal quench is induced within approximately $5~\mathrm{ms}$, and the plasma subsequently enters the current quench phase after about $13~\mathrm{ms}$. These findings confirm that the integrated system can establish an effective and coordinated chain of data acquisition, prediction, and mitigation on EXL-50U.

Due to real-time constraints, the current model utilizes only eight diagnostic channels as inputs, and has not yet incorporated richer physical information such as magnetic probe arrays, flux loops, equilibrium reconstruction, or locked mode signals. In future work, additional real-time diagnostics and magnetohydrodynamic (MHD)-related features will be integrated. Furthermore, the warning criteria and deployment strategy will be further optimized to enhance both the prediction performance and the engineering applicability of the system.

\section{Acknowledgments}\label{sec:acknowledgments}

This work is supported by National Natural Science Foundation of China under Grant No.12275142 and No.12275354, National MCF Energy R\&D Program under Grant No.2024YFE03020001. Besides, we are particularly grateful to ENN for their support of this work.

\end{document}